\documentclass[conference]{IEEEtran}
\IEEEoverridecommandlockouts
\usepackage{cite}
\usepackage{amsmath,amssymb,amsfonts}
\usepackage{algorithm}
\usepackage{algorithm}
\usepackage{algpseudocode}
\usepackage{graphicx}
\usepackage{textcomp}
\usepackage{xcolor}
\usepackage{bm}
\usepackage{multicol}
\usepackage{listings}
\usepackage[left=1.02in,right=1.02in,top=0.75in, bottom=1.04in]{geometry}
\algnewcommand{\IIf}[1]{\State\algorithmicif\ #1\ \algorithmicthen}
\algnewcommand{\ElseIIf}[1]{\algorithmicelse\ #1} 
\algnewcommand{\EndIIf}{\unskip\ \algorithmicend\ \algorithmicif}
\def\BibTeX{{\rm B\kern-.05em{\sc i\kern-.025em b}\kern-.08em
    T\kern-.1667em\lower.7ex\hbox{E}\kern-.125emX}}
\begin{document}

\newtheorem{example}{Example}

\title{Parallelizing the stabilizer formalism for quantum machine learning applications\\

\thanks{}
}

\author{
	\IEEEauthorblockN{Vu Tuan Hai,  Le Vu Trung Duong, Pham Hoai Luan, and Yasuhiko Nakashima}
	\IEEEauthorblockA{
    Nara Institute of Science and Technology, 8916–5 Takayama-cho, Ikoma, Nara 630-0192, Japan.\\
Email: vu.tuan\_hai.vr7@naist.ac.jp} 
}

\maketitle

\begin{abstract}

The quantum machine learning model is emerging as a new model that merges quantum computing and machine learning. Simulating very deep quantum machine learning models requires a lot of resources, increasing exponentially based on the number of qubits and polynomially based on the depth value. Almost all related works use state-vector-based simulators due to their parallelization and scalability. Extended stabilizer formalism simulators solve the same problem with fewer computations because they act on stabilizers rather than long vectors. However, the gate application sequential property leads to less popularity and poor performance. In this work, we parallelize the process, making it feasible to deploy on multi-core devices. The results show that the proposal implementation on Python is faster than Qiskit, the current fastest simulator, 4.23 times in the case of 4-qubits, 60,2K gates.

\end{abstract}


\section{Introduction}

Parameterized Quantum Circuit (PQC) nowadays is a popular model for Quantum Machine Learning (QML) models \cite{quantumcircuitlearning}, A PQC $U(\bm\theta)$ is composed of $m$ quantum gates $g_j$, which include fixed and parameterized gates. Parameterized gates take a role as learning nodes in classical machine learning models, which hold a trainable parameter, ranging from $0$ to $2\pi$. These parameters can be updated via the Parameter-Shift Rule (PSR) technique, which provides an analytic gradient through $2m$ quantum evaluations \cite{Wierichs2022generalparameter}. PQCs are required for some QML models such as Variational Quantum Eigensolver \cite{TILLY20221}, Quantum Approximate Optimization Algorithm \cite{BLEKOS20241} and Quantum Neural Network \cite{PRXQuantum.5.020328}.

Due to the wide range of simulation approaches, many quantum simulators have been proposed, such as state-vector \cite{javadiabhari2024quantumcomputingqiskit}, tensor-network \cite{10313722}, decision diagram \cite{Vinkhuijzen2023limdddecision} and stabilizer formalism \cite{Gidney2021stimfaststabilizer}. The goals of quantum simulator development range from domain-specific to general purpose, then reach the boundary of quantum advantage where classical computers can no longer simulate a quantum system in acceptable runtime. However, almost all simulators try to simulate the PQC efficiently. The scalability of the simulator is evaluated by the number of qubits  $n$ (\#Qubits) and the number of gate applications $m$. Most simply, in the state-vector approach, the quantum state $|\psi(\bm\theta)\rangle$ is a state vector that can be represented as a $2^n$ - dimensional complex tensor. As a result, the difficulty when making a quantum simulator is the exponential raising of both execution time and memory space based on \#Qubits. While the number of gates only increases the execution time linearly. Because the exponential factor is unbreakable due to the quantum system property, recent works related to software techniques focus on how to reduce the gate application time.

Stabilizer formalism based on Heisenberg's picture provides a compact presentation for gate application. The action of gates on the stabilizer is that mapping between $n$ - qubits Pauli string $\mathcal{P}_n$ rather than matrix-vector multiplication and conducting $g_j|\psi\rangle$ in stabilizer formalism will be simpler than the state-vector approach; then it's an advantage if we perform many gate circuits by stabilizers. Existing stabilizer formalism packages such as Qiskit \cite{javadiabhari2024quantumcomputingqiskit} (Rust), Cirq \cite{isakov2021simulationsquantumcircuitsapproximate} (Python), Stim \cite{Gidney2021stimfaststabilizer} (C++), and PyClifford  (Python) have been developed and used widely. Although the stabilizer formalism operates on Clifford gates, which can run fast in a hundred qubits, it limits the application of the quantum simulator, especially for PQC. On the other hand, if we use non-Clifford gates (extended stabilizer formalism), such as $R_j(\theta) = \{R_x(\theta), R_y(\theta), R_z(\theta)\}$, it leads to a faster increase in complexity than the state-vector approach based on $\#$Qubits. As shown in \cite{modelcounting}, the author shows the exponential raising of stabilizer order in cases of QNN, which require non-Clifford gates. The other research solves this problem by approximation method, reducing the complexity but also decreasing the accuracy \cite{Bravyi2019simulationofquantum}. Another weakness of stabilizer formalism is the sequential algorithm, which means running the software packages on GPUs is insufficient. 

As mentioned above, there are two use cases of stabilizer formalism: the first is for large-qubit Clifford circuits, and the second is the small-qubit and deep general circuits which require thousands of both Clifford and non-Clifford gates. In this research, \textbf{we focus on the second use case}. While Clifford circuits are only used for quantum error correction and benchmarking the performance of quantum hardware. \textbf{The deep general circuits used for deep PQC cover a wide range of upcoming applications}. Our proposed algorithm - PStabilizer (Parallel stabilizer formalism) covers the big weakness of the original algorithm. Instead of executing $g_j|\psi\rangle$ gate by gate, our algorithm performs the action of the operator $U$ ($U|\psi\rangle$) by grouping the same sequential gate type. $U|\psi\rangle$ is then being parallelized, making it feasible to apply speed-up techniques.  



\section{Background}
\label{sec:background}

\subsection{Stabilizer formalism}

A $n$-qubit quantum state $|\psi\rangle$ is defined as a stabilizer state by a unitary operator $U$ if and only if it is a $+\mathbb{1}$ eigenvector of $U$, mean $U|\psi\rangle=|\psi\rangle$. The stabilizer states form a strict subset of all quantum states which can be uniquely described by maximal commutative subgroups of the Pauli group, called the stabilizer group. The elements of the stabilizer group are called stabilizers, which are represented by the sum of the weighted Pauli string. $\mathbb{P}_n=\sum_j  \lambda_j P_{n,j},\;\lambda P_n=\lambda p_0\otimes \ldots \otimes p_{n-1},\;$ where $\lambda\in\mathbb{R}$ is the weight,  $P_n$ is the Pauli string which is composed from $n$ Pauli matrices $p_j$, $p_j\in \{I,X,Y,Z\}$. Recall that the Clifford group is formed by unitary operators mapping the Pauli group to itself. Any stabilizer group $G$ can be specified by a set of stabilizers so that every element inside $G$ can be obtained through Matrix Multiplication (MM) between  $\mathbb{P}_{n, i}, \mathbb{P}_{n, j}$, and denoted $\langle \mathbb{P}\rangle = G$. These stabilizers are initialized and transformed by $m$ gates as~\eqref{eq:generator}:

\begin{align}
    G=\left\langle\begin{array}{c}
    \mathbb{P}_{n, 0}^{(0)} \\
    \vdots \\
     \mathbb{P}_{n, n-1}^{(0)} \\
    \end{array}\right\rangle\xrightarrow{g^{(m)}}\left\langle\begin{array}{c}
    \mathbb{P}_{n, 0}^{(m)} \\
    \vdots \\
     \mathbb{P}_{n, n-1}^{(m)} \\
    \end{array}\right\rangle,\label{eq:generator}
\end{align}

where $\mathbb{P}_{n, j}^{(0)}\equiv\mathbb{Z}_{n,j}=I^{\otimes j}\otimes Z\otimes I^{\otimes (n-j-1)}$. We can relate the (generators of the) stabilizer group directly to the stabilizer state $|\psi\rangle$, as the target density operator can be obtained through the product between stabilizers, or a sum of Pauli matrices as ~\eqref{eq:density_operator1} and ~\eqref{eq:density_operator2}, respectively:

\begin{align}
    \rho\equiv|\psi\rangle\langle\psi|&=\frac{1}{2^n}\prod_{j=1}^{n-1} \left(I^{\otimes n}+\mathbb{P}^{(T)}_{n,j}\right) \label{eq:density_operator1}\\
    &=\frac{1}{2^n} \sum_{\mathcal{P}_n \in \mathcal{P}_{|\psi\rangle}} \lambda_{\mathcal{P}_n}\mathcal{P}_n,
    \label{eq:density_operator2}
\end{align}

It can be used to get some physical properties such as measurement value by computing $p_{j,k} = \text{tr}([\frac{1}{2}(\mathbb{I}+Z_k)]|\psi\rangle\langle\psi|)$, or simply track on $\rho_{0,0}$. If a Clifford gate $g$ is applied to the stabilizer state and let $\mathbb{P}_n \in {G}$, then $U \mathbb{P}_n U^{\dagger} g|\psi\rangle=g|\psi\rangle$. To be specific, acting a Clifford gate on $j^{\text{th}}$ qubit (notated as $g_j$), to a stabilizer $\mathbb{P}_n=\lambda\;p_0 \otimes \ldots \otimes p_{n-1}$ return $g_j \mathbb{P}_n g_j^{\dagger}= \lambda\;p_0 \otimes \ldots \otimes g_j p_j g_j^{\dagger} \otimes \ldots \otimes p_{n-1}$. Since $g_j$ is a Clifford gate and only the $j^{\text{th}}$ entry needs to be updated; reducing $g_jS g_j^{\dagger}$ can be done in constant time.


\subsection{Limitation of stabilizer formalism}

\begin{figure}[ht]
    \centering
    \includegraphics[width=0.95\linewidth]{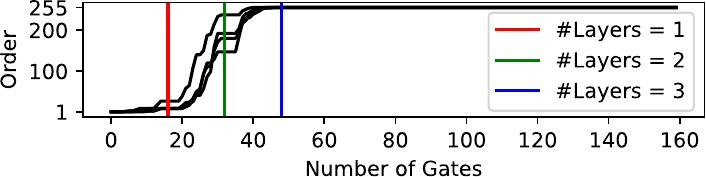}
    \vspace{-0.3cm}
    \caption{The stabilizer order is based on the number of gates (follow $W_{\text{chain}}+ZXZ$ ansatz). The gate application's order is from qubit $0^{\text{th}}$ to qubit $(n-1)^{\text{th}}$, left to right. We are considering the worst cases.}
    \label{fig:stabilizer_order}
\end{figure}

The act of non-Clifford gates turns a single Pauli matrix into a Pauli term. If these gates act on every qubit, a stabilizer, initial as a Pauli string easily turns into the product of $n'$ Pauli strings, note that $|\mathbb{P}_{n}|=n'$ called stabilizer order. $n'$ can be $3^n$, which can increase to maximally $4^n$ after CX actions. The next actions require a loop through all strings, making it much slower than the state vector which only traverses on $2^n$ - complex entries. Fig.~\ref{fig:stabilizer_order} shows how stabilizers are becoming more complex through non-Clifford and CX gate applications.

\begin{example}
    $\mathbb{P}_{1}=X\xrightarrow{R_y(\theta_1)}\cos(\theta_1)X-\sin(\theta_1)Z$\\$
    \xrightarrow{R_x(\theta_2)} \cos(\theta_1)X-\sin(\theta_1)\cos(\theta_2)Z+\sin(\theta_1)\sin(\theta_2)Y$
    \label{example:mapped_pauli}
\end{example}

\section{Proposed techniques}
\label{sec:proposed}
\begin{figure}[ht]
    \centering
    \includegraphics[width=0.99\linewidth]{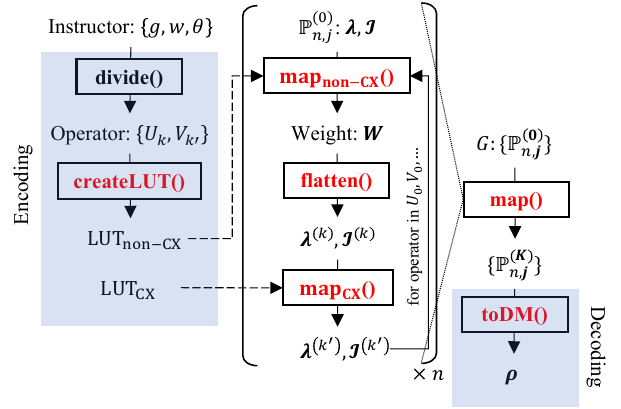}
    \vspace{-0.6cm}
    \caption{Architecture of PStabilizer algorithm. All the \textcolor{red}{red} functions are parallelizable. The arrow stands for data flow, dotted arrows are notated for one-time loading. }
    \label{fig:pipeline}
\end{figure}

For deep circuits, the main execution time comes from applying gates on stabilizers, since this function transforms $j^{\text{th}}$ Pauli matrix in every Pauli string, the time complexity for extended stabilizer formalism is $\mathcal{O}(mn4^n)$, $m$ is the number of gates. The following technique reduces the complexity to $\mathcal{O}((K+K')n4^n)$, $K,K'\ll m$. Because each stabilizer is independent, we only describe the technique applying to a single stabilizer; for the whole generator, this process can be run $n$ times parallel. The architecture of PStabilizer follows the Encode-Decode, the Encoding stage converts the stabilizer to tensor format, and the tensor is converted back to stabilizer at the Decoding stage. 

\subsection{Dividing circuit}

\begin{figure}
    \centering
    \includegraphics[width=1.02\linewidth]{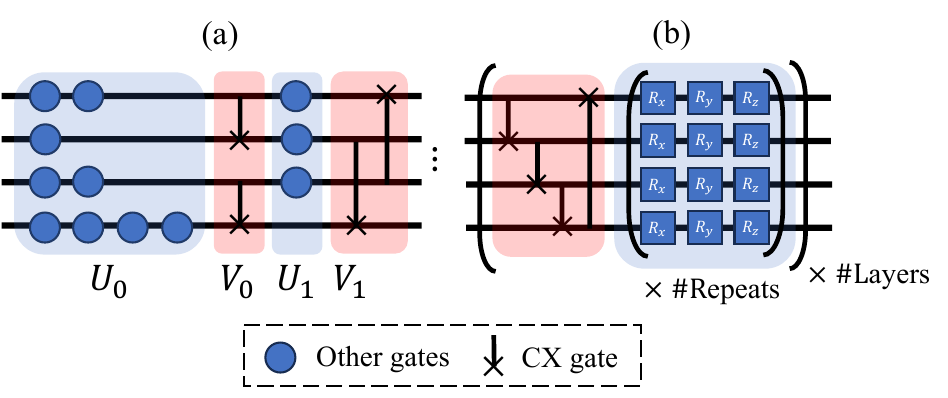}
    \vspace{-0.8cm}
    \caption{(a) A quantum circuit can be divided into $\{U_k\},\{V_k\}$, and end up with $U_{K-1}$ or $V_{K'-1}$ (b) $W_{\text{chain}}+ZXZ$ topology used for experiments.}
    \label{fig:operator}
\end{figure}

We present gates as instructors (same as Qiskit \cite{javadiabhari2024quantumcomputingqiskit}). An instructor is composed as the tuple $\{g, w, \theta\}$, present for gate name, qubit (wire), and parameter value, respectively. In case $g\notin R_j$, $\theta$ is set as $0$.  Performing gate-by-gate on stabilizers leads to slow and delayed computations. Hence, we consider performing operators instead, each operator includes the same type of gates. Because the CX gate has different behavior from other gates, a quantum circuit is divided into $K$ $U_k$ (where each $U_k$ has only non-CX gates) and $K'$ $V_k$ (where each $V_k$ has only CX gates). These operators interleaved each other, as presented in Fig.~\ref{fig:operator}, obey $|K-K'|\leq 1$. Next, we group instructors inside each operator based on its wire, notated $U_{k,j}/V_{k,j}$ for instructors in the operator $k^{\text{th}}$ and at wire $j^{\text{th}}$.

\subsection{Encoding}

A basic stabilizer is presented as $\bm\lambda$ and corresponding index vector $\mathcal{I}$ where each index $i\in\mathcal{I}$ is encoded from Pauli string using base-$4$ $(I=0,X=1,Y=2,Z=3)$. An index ranges from $0$ ($I^{\otimes n}$) to $4^n-1$ ($Z^{\otimes n}$).

\begin{example}
$XYZ\xrightarrow{\text{stringToIndex()}}1\times4^2+2\times4+3=27$, on the other hand, it requires the number of qubits to convert an index to a Pauli string ($\text{indexToString()}$).
\label{example:index_to_word}
\end{example}

A single Pauli matrix in the Pauli string can be mapped into a sum of weighted Pauli matrices $P=w_0I+w_1X+w_2Y+w_3Z$ as Example.~\ref{example:expand_reduce}. Generally, a mapped stabilizer is:

\begin{align}
\mathbb{P}_{n,j}^{(t)}=\sum_{j=0}^{n'-1}\lambda_j\left(\prod_{k=0}^{n-1} P_{k}\right),
\end{align}


which need to be expanded and reduced to the basic form. Simplicity, any sum of the weighted Pauli matrix can be presented as $\bm{w}\equiv[w_0,w_1,w_2,w_3]$. In the worst case ($n'=4^n$), a mapped stabilizer is equivalent to $4^n\times n\times 4$ - tensor. The expand and reduce operation on this tensor can also parallelize.

\begin{example}
    $IX + (X + Y)Y$\\
    $\xrightarrow{\text{encode}()} [[[1,0,0,0],[0,1,0,0]],[[0,1,1,0],[0,0,1,0]]]$\\$\xrightarrow{\text{expand}()/\text{reduce}()}[0,1,0,0,0,0,1,0,0,0,1,\ldots,0]$\\
    $\equiv IX+XY+YY$.
    \label{example:expand_reduce}
\end{example}

\subsection{Gate}

The mapped stabilizer after $m$ gates, specifically on multiple Pauli strings $\{\mathcal{P}\}$, is conservative because of the fixed size of the encoded weight. We consider two types of gates: non-CX (including $H, S, R_j$) and CX gate. It would be efficient if we do $m$ non-CX gates at a single step (in one $U_k$) without expanding and reducing. This function is notated as $\text{map}_{\text{non-CX}}: \bm{w}\times g\rightarrow \bm{w}$ which can be applied $m$ times sequentially as~\eqref{eq:map_noncx}:

\begin{align}
    \{X,Y,Z\}&\xrightarrow{\text{PauliToIndex}()}\text{index}\nonumber\\
    &\xrightarrow{\text{map}^{(m)}_{\text{non-CX}}(\ldots, U_{k,j})}P^{(t+m)}.
    \label{eq:map_noncx}
\end{align}


The action of the gate follows up a rule, conducting $\text{map}_{\text{non-CX}}(\ldots,g_j)$ is tracking through Tab.~\ref{tab:mapper_noncx}. Because the mapping function only acts on $X, Y$ or $Z$, it is efficient if we construct a look-up table called $\text{LUT}_{\text{non-CX}}$ to track the output of every $\text{map}^{(m)}_{\text{non-CX}}(\ldots, U_{k,j})$ function. If $|\mathbb{P}_n|=4^n$, this reduces $(2\times 4^n)/3$ duplicate computation times.

A CX gate acts on two indices $i,j$ which change the Pauli $i,j$ in the Pauli string to another $(XI\leftrightarrow XX, IY\leftrightarrow ZY, YI\leftrightarrow YX, IZ\leftrightarrow ZZ, XY\leftrightarrow YZ)$ or unchanged $(II, IX, ZI, ZX)$, or change both Pauli and sign $(XZ\leftrightarrow -YY)$. For $m'$ CX gate, the array $\mathcal{I}$ is suffered $m'$ times. Because $i,j\in[0,n-1];i\neq j$, there are only $n(n-1)$ different CX gate for $n$ qubits. The size of $\text{LUT}_{\text{CX}}$ will be $n\times (n-1) \times 4^n$, the value at each index (index) in an array $(\text{LUT}_{\text{CX}})_{i,j}$ is the output of $\text{map}_{\text{CX}}(\text{index},i,j)$.

\begin{figure}
    \centering
    \includegraphics[width=0.99\linewidth]{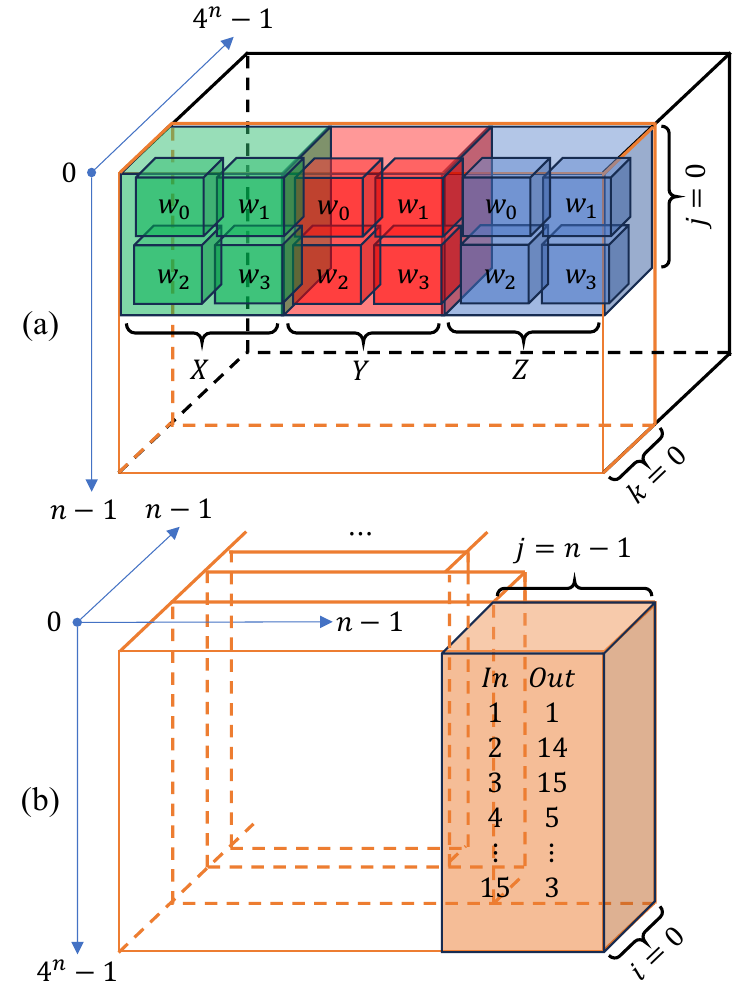}
    \vspace{-0.8cm}
    \caption{LUT for (a) $U_k$ as $K\times n\times3\times 4$ - tensor and (b) $V_k$ as $n\times (n-1)\times 4^n$ - tensor.}
    \label{fig:lut}
\end{figure}

\begin{table}[ht]
    \caption{Output for each 1-qubit gate on $\mathcal{P}$ with input $\bm w=[w_0,w_1,w_2,w_3]$.}
    \centering
    \vspace{-0.2cm}
    \resizebox{0.49\textwidth}{!}{
    \begin{tabular}{|c|c|}
        \hline
         Gate& Output\\\hline
         $H$& $[w_0, w_3, -w_2, w_1]$\\\hline
         $S$&$[w_0, -w_2, w_1, w_3]$ \\\hline
         $R_x(\theta)$& $[w_0, w_1, w_2 \cos(\theta) - w_3\sin(\theta), w_2\sin(\theta) + w_3\cos(\theta)]$ \\\hline
         $R_y(\theta)$&$[w_0, w_1\cos(\theta) + w_3\sin(\theta), w_2, w_3\cos(\theta) - w_1\sin(\theta)]$ \\\hline
         $R_z(\theta)$& $[w_0, w_1\cos(\theta) - w_2\sin(\theta), w_2\cos(\theta) + w_1\sin(\theta), w_3]$ \\\hline
    \end{tabular}}
    \vspace{-0.5cm}
    \label{tab:mapper_noncx}
\end{table}


\subsection{Decoding}

After being transformed by gates, $\{\bm\lambda^{(K)}, \mathcal{I}^{(K)}\}$ is decoded as sparse matrices. For each pair $\{\lambda, \text{index}\}$:

\begin{align}
    \lambda, \text{index}\xrightarrow{\text{indexToString}()} \lambda \mathcal{P}_n\xrightarrow{\text{toMatrix()}}\{\text{col},\text{val}\},
\end{align}

where $\{$col, val$\}$ are the column indices and values for Compressed Sparse Row (CSR) format \cite{10.1145/1583991.1584053}, respectively. $\lambda \mathcal{P}_n$ is the PauliComposer or PauliDiagonalComposer object in case $p_j \in\{X,Z\},\;\forall p_j\in\mathcal{P}_n$\cite{VidalRomero2023}.

Normally, this operation took $n-1$ MM between $2^n\times2^n$ matrices. However, these matrices are limited to Pauli matrices, opening up the chance for applying sparse optimization techniques. The first notice is that we only use binary matrices; to deal with the imaginary part in $Y$, 
 we transform $Y\rightarrow i\tilde{Y}$, then $\lambda \mathcal{P}_n\rightarrow(\lambda i^{n_{Y}\;\text{mod}\;4})\tilde{\mathcal{P}}_n$. Next, a key property of Pauli strings is that on each row and column, it has only one nonzero entry (which has value $\pm1$), because the same property comes from $\{I, X, Y, Z\}$. It realizes that computing the sparse form of any Pauli string turns to find the location and sign of nonzero entries, instead of performing $n-1$ tensor multiplication. The algorithm proposed by~\cite{VidalRomero2023} has completed this task by looping through all rows, done in $\mathcal{O}(2^n)$ addition and $\mathcal{O}(2^n)$ changes of sign for worst cases, much smaller than $\mathcal{O}(n2^{2n})$ or $O(n2^n)$ additional/multiplication for dense or sparse MM, respectively.

\section{Parallelization of stabilizer formalism}
\label{sec:algorithm}

\begin{algorithm}[ht]
\caption{constructLUT()} 
\label{algo:pstabilizer}
\begin{algorithmic}[]
\Require $\{\{U_{k,j}\}\}$ and $\{V_k\}$
\State $\text{LUT}_{\text{non-CX}}\gets \bm 0^{K\times n\times3\times4}$
\State $\text{LUT}_{\text{CX}}\gets \bm 0^{n \times (n-1) \times 4^n}$
\For{$k$ in $[0\ldots K-1]$ in parallel}
    \For{$j$ in $[0\ldots n-1]$ in parallel}
        \State $(\text{LUT}_{\text{non-CX}})_{k,j,0}\gets\text{map}_{\text{non-CX}}(X,U_{k,j})$
        \State $(\text{LUT}_{\text{non-CX}})_{k,j,1}\gets\text{map}_{\text{non-CX}}(Y,U_{k,j})$
        \State $(\text{LUT}_{\text{non-CX}})_{k,j,2}\gets\text{map}_{\text{non-CX}}(Z,U_{k,j})$
     \EndFor    
\EndFor

\For{$k$ in $[0\ldots n-1]$ in parallel}
    \For{$j$ in $[0\ldots n-1]$ in parallel}
        \If{$i\neq j$}
            \For{$i$ in $[0\ldots4^n-1]$}
                \State $(\text{LUT}_{\text{CX}})_{k,j,i}\gets\text{map}_{\text{CX}}(i, j, k, n)$
            \EndFor
        \EndIf
     \EndFor    
\EndFor
\State \Return $\text{LUT}_{\text{nonCX}},\text{LUT}_{\text{CX}}$
\end{algorithmic}
\end{algorithm}

\subsection{Construct look-up table}

We called $\text{LUT}_{\text{non-CX}}$ and $\text{LUT}_{\text{CX}}$ are ExtTableaus (extended stabilizer tableaus \cite{Bravyi2019simulationofquantum}) for both Clifford and non-Clifford circuits. As shown in Algorithm.~\ref{algo:pstabilizer}, each $\text{map}_{\text{non-CX/CX}}(\ldots)$ runs independently. Constructing the $\text{LUT}_{\text{non-CX}}$ and $\text{LUT}_{\text{CX}}$ can accelerate $K\times n\times 3$ and $n\times(n-1)\times 4^n$ times, respectively; depend on the maximum number of cores.

\subsection{Map}

The order of operators in the Map stage is decided based on the first instructor. If the first instructor's name is CX, the order will be $V_0, U_0, V_1, U_1,\ldots$ and vice versa. In the PStabilizer, Map is the only sequential stage, the output of any operator is the input of the next operator, and so on until the final operator. The bottleneck comes from $\text{flatten}()$ and $\text{map}_{\text{CX}}()$ which are two highly complex functions.

Because CX-gates require the basic form of stabilizer, the mapped stabilizer must be transformed from a sum of the product of sum (present by a $n' \times n \times 4$ - tensor) to a sum ($\bm\lambda^{(t+1)}$), as~\eqref{eq:flatten}:

\begin{align}
    \bm\lambda^{(t+1)} \xleftarrow{\text{flatten}()} \sum_{j=0}^{n'-1}\left\{\prod_{k=1}^{4^n}\left(\bigotimes_{l=1}^{n}\bm W^{(t)}_{j,l}    \right) \right\},
    \label{eq:flatten}
\end{align}

where $\bigotimes$ is Cartesian product notation. This function computes the product of $4^n$ combination of every $n\times 4$ - weight matrix, and then sums over columns to get $\bm\lambda^{(t+1)}$. Note that $I=[1,0,0,0]$ is conservative for any $U_j$, the number of combinations can be reduced $(3/4)^{n_I}$ for Pauli string $P$ hold $n_I$ Pauli $I$.

\begin{example}
    $\mathbb{P}=(I+2X+3Y+4Z)(I+2X+3Y+4Z) + (I+2X+3Y+4Z)(I+2X+3Y+4Z)$\\$\xrightarrow{\text{encode}()} [[[1,2,3,4], [1,2,3,4]], [[1,2,3,4], [1,2,3,4]]]$ is the weights for the 2-qubit system with $\overline{n}=2$, $\bm{W}\xrightarrow{\text{flatten()}}\mathbb{R}^{n'\times 4^2}=
    \begin{bmatrix}
        1\;2\;\ldots\;16\\
        1\;2\;\ldots\;16  
    \end{bmatrix}\xrightarrow{\text{reduce}()}[2\;4\;\ldots\;32]$. 
\end{example}

\begin{figure}[ht]
    \centering
    \includegraphics[width=0.95\linewidth]{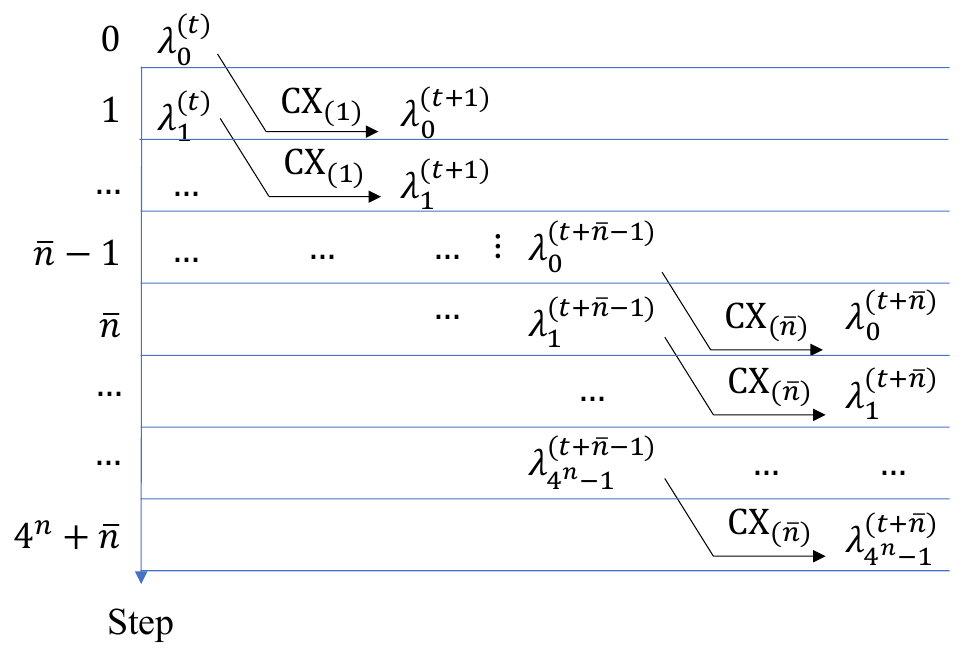}
    \vspace{-0.5cm}
    \caption{Pipeline architecture for $\overline{n}$ CX gates act on the array $\bm\lambda^{(t)}$.}
    \label{fig:map_cx}
\end{figure}


\begin{figure}[ht]
    \centering
    \includegraphics[width=0.99\linewidth]{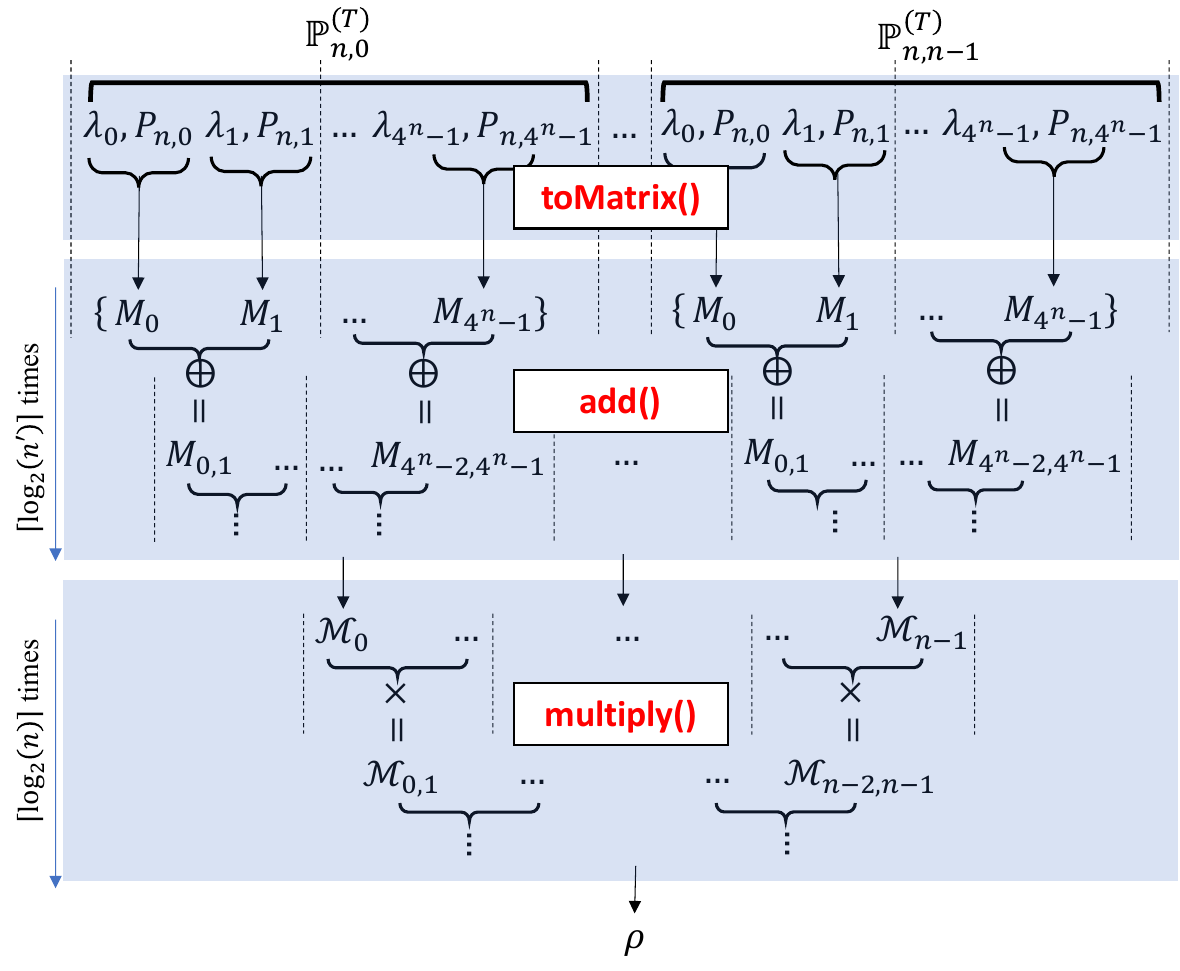}
    \vspace{-0.8cm}
    \caption{Decoding stage, includes three functions: toMatrix(), add(), and multiply().}
    \label{fig:toDM}
\end{figure}

\subsection{Density matrix}

A stabilizer needs to convert to a $2^n\times 2^n$ - matrix:

\begin{align}
    \mathbb{I}^{\otimes n}+\sum_{j=1}^{n^{'}-1}\lambda_j\mathcal{P}_n\xrightarrow{\text{add()}}\mathcal{M}
\end{align}


using divide-and-conquer strategy \cite{10.1145/3432261.3432271} as Fig.~\ref{fig:toDM}. In the worst case, this task can be done in $\log_2(4^n)= 2n$ steps. The next stage involves the $n$ stabilizers, all matrices $\{\mathcal{M}_j\}$ conduct a chain multiplied in $\log_2(n)$ times by the same strategy, returning density matrix $\rho$.

\section{Experiments}
\label{sec:experiments}
\begin{figure*}[ht]
    \centering
    \includegraphics[width=0.99\linewidth]{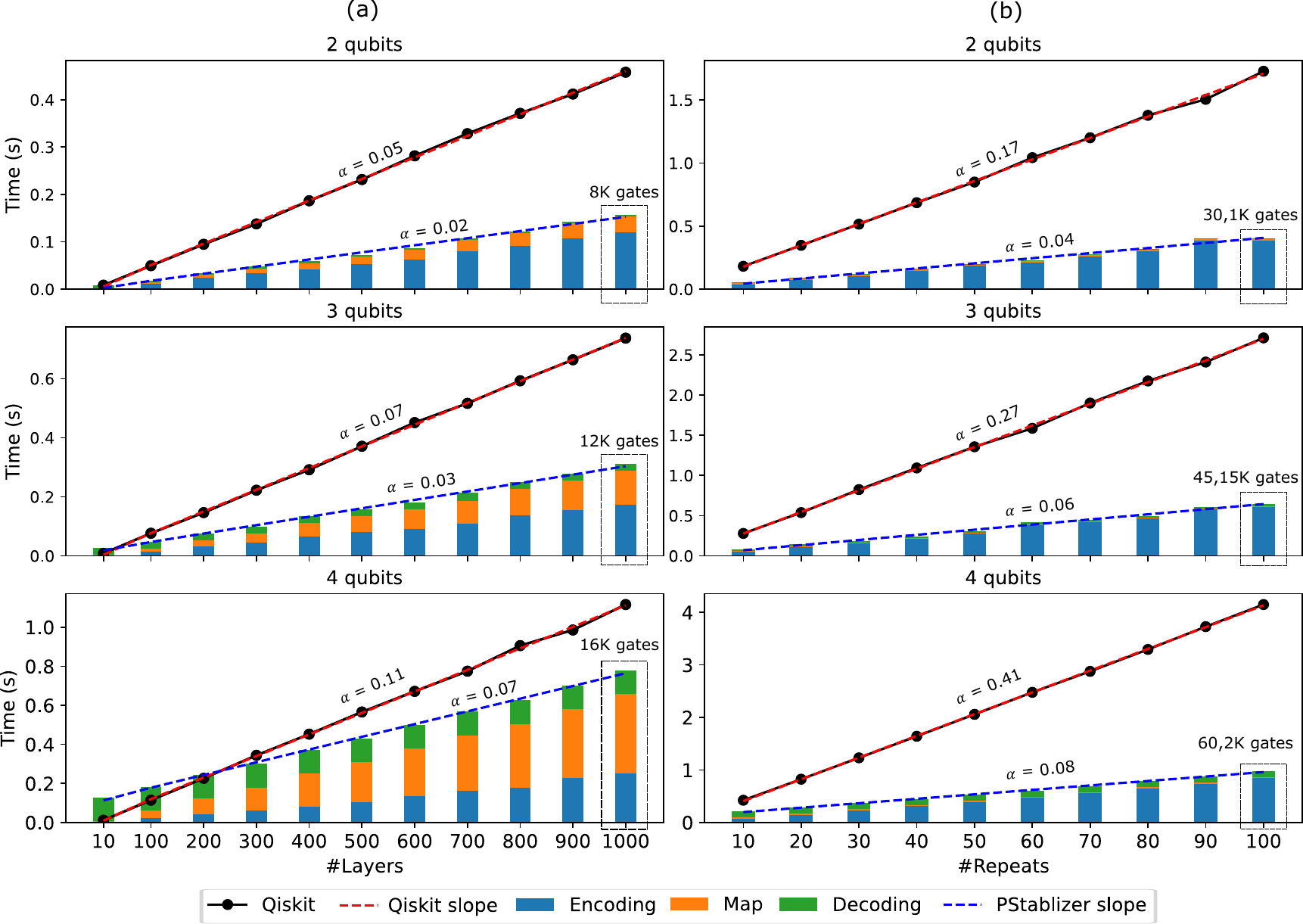}
    \vspace{-0.2cm}
    \caption{Execution time of PStabilizer (sum of Encoding, Map, and Decoding) and Qiskit on various (a) $\#$Layers and (b) $\#$Repeats.}
    \label{fig:times}
\end{figure*}

The PStabilizer algorithm is implemented in Python 3.11 version with Jax \cite{jax2018github} as a supporter. The comparable software is Qiskit version 1.1.1 \cite{javadiabhari2024quantumcomputingqiskit}, implemented by Python interface and Rust core. All experiments are running on Intel(R) i9-10940X CPU @ 3.30GHz. We evaluate the task $\text{toDM()}$ for both packages, the execution time is measured from initializing the stabilizers until we get the final density matrix. All cases run at least $10$ times, then take the average. Because of the limited number of process cores, the Map stage is measured on one stabilizer, then assumes that if it can deployed on multiple devices, the execution time will be the same. Other parallelizable functions run in the sequential mode. 
The codes used for this study are available upon reasonable request.

The benchmarked ansatz is $|W_\text{chain}+ZXZ\rangle$ with random parameters on the ZXZ part (rotation part). The overall ansatz and ZXZ part are duplicated $\#\text{Layers}$ times and $\#\text{Repeats}$ for investigating the scalability of PStabilizer. Because this simulator focuses on low qubits, but deep circuits, the simulated qubit ranges only from 2 to 4, but the $\#\text{Layers}$ and $\#\text{Repeats}$ are up to $1000$ (with $\#\text{Repeats}=1$) and $100$ (with $\#\text{Layers}=50$). Note that in this case, $K=K'=\#\text{Layers}$.

PStabilizer shows the weakness scale on $\#\text{Layers}$, as the number of qubits increased, from $\textbf{2.9}$ times faster ($0.157$ vs $0.4578$) at $2$ qubits to $\textbf{1.43}$ times faster ($0.7789$ vs $1.1158$) at $4$ qubits. Although the slope is smaller than Qiskit, PStabilizer will be outperformed in the next few qubits. As predicted, the longest stage belongs to the Map stage, which increases linearly based on $\#$Layers. In Fig.~\ref{fig:times} (b), PStabilizer shows superiority over Qiskit in case of the enormous number of gates, but small $K/K'$. Because small $\#$Layers, the Map stage consumes a little time compared with the Encoding stage. Furthermore, Qiskit increases faster based on the number of gates.

\section{Conclusion}
\label{sec:conclusion}

The PStabilizer is the parallel version of stabilizer formalism using encoding-decoding architecture. While operating on multi-dimensional tensor speed-up by both software and hardware rather than original stabilizers, PStabilizer can take advantage of multiple core processors. This approach offers a short execution time in case of a thousand gates, faster than Qiskit, the fastest quantum simulation package implemented by Rust, and even PStabilizer is implemented by Python. Note that the complexity now depends on the number of operators, so the PStabilizer is suitable for structured circuits and does not achieve the best for random circuits.

However, notice that PStabilizer only achieves the best performance on low $\#$Qubits and structured circuits, or low $K/K'$; usable for quantum machine learning models. For higher $\#$Qubits or random circuits, PStabilizer will be slower than other types of simulator due to the exponential rise of stabilizer order, $\mathcal{O}(4^n)$ compared to $\mathcal{O}(2^n)$ from the state-vector approach. The next versions focus on optimizing constructLUT() and flatten(), which is the bottleneck of PStabilizer. Note that in this work, PStabilizer is just deployed on the CPU, therefore implementing it on efficient hardware such as FPGA/CGRA/GPU will also be considered.




\section*{Acknowledgment}

This work was supported by JST-ALCA-Next Program Grant Number JPMJAN23F4, Japan, and the Next Generation Researchers Challenging Research Program under number zk24010019. The research has been partly executed in response to the support of JSPS, KAKENHI Grant No. 22H00515, Japan. 



\bibliographystyle{IEEEtran}
\bibliography{ref.bib}

\end{document}